# Modeling Dabrafenib Response Using Multi-Omics Modality Fusion and Protein Network Embeddings Based on Graph Convolutional Networks

La Ode Aman[1*], A Mu'thi Andy Suryadi[1], Dizky Ramadani Putri Papeo[1], Hamsidar Hasan[1], Ariani H Hutuba[1], Netty Ino Ischak[2], Yuszda K. Salimi[2]

[1] Department of Pharmacy, Faculty of Sports and Health, Universitas Negeri Gorontalo, Jl. Jend. Sudirman No.6, Dulalowo Timur., Kec. Kota Tengah, Gorontalo, 96128, Indonesia
[2] Department of Chemistry, Mathematics and Natural Sciences, Universitas Negeri Gorontalo, Jl. Prof. Dr. Ing. B. J. Habibie, Kecamatan Tilongkabila, Kabupaten Bone Bolango, Gorontalo 96119, Indonesia

*Corresponding author: laode_aman@ung.ac.id

## ABSTRACT

Cancer cell response to targeted therapy arises from complex molecular interactions, making single omics insufficient for accurate prediction. This study develops a model to predict Dabrafenib sensitivity by integrating multiple omics layers (genomics, transcriptomics, proteomics, epigenomics, and metabolomics) with protein network embeddings generated using Graph Convolutional Networks (GCN). Each modality is encoded into low dimensional representations through neural network preprocessing. Protein interaction information from STRING is incorporated using GCN to capture biological topology. An attention based fusion mechanism assigns adaptive weights to each modality according to its relevance. Using GDSC cancer cell line data, the model shows that selective integration of two modalities, especially proteomics and transcriptomics, achieves the best test performance ($R^2$ around 0.96), outperforming all single omics and full multimodal settings. Genomic and epigenomic data were less informative, while proteomic and transcriptomic layers provided stronger phenotypic signals related to MAPK inhibitor activity. These results show that attention guided multi omics fusion combined with GCN improves drug response prediction and reveals complementary molecular determinants of Dabrafenib sensitivity. The approach offers a promising computational framework for precision oncology and predictive modeling of targeted therapies.

# INTRODUCTION

Resistance and variable response to anticancer drugs remain major challenges in precision oncology. Molecular heterogeneity—encompassing genomic, transcriptomic, proteomic, epigenomic, and metabolomic variations—causes cancer cells from different patients to exhibit distinct therapeutic responses even when given the same treatment (Gouru et al., 2025). Therefore, multi-omics integration has become an essential strategy for understanding biological complexity and improving the accuracy of drug sensitivity prediction.

However, multi-omics integration faces several obstacles: the data are extremely high-dimensional, heterogeneous (each modality has its own characteristics), and often redundant or noisy. Traditional methods such as linear regression or matrix factorization often fail to capture the nonlinear interactions and biological complexity across molecular layers (Zhang et al., 2025; Partin et al., 2023). For this reason, modern computational approaches based on deep learning and graph neural networks (GNNs) are increasingly popular due to their ability to model complex and structured data.

One promising approach is to leverage biological interaction networks—such as protein–protein interaction (PPI) networks—together with omics data to capture functional relationships between genes/proteins. By mapping gene expression or other omics features to nodes in the network and applying Graph Convolutional Networks (GCNs), the model can learn representations that incorporate biological topology and functional correlations among proteins. Early studies such as the DrugGCN framework have shown that combining gene expression with PPI networks can improve drug-response prediction accuracy compared to genome-only methods (Kim et al., 2021).

Furthermore, multi-omics integration using fusion and network-based approaches has been shown to enhance predictive performance. For example, the MOMLIN model integrates multiple omics modalities and demonstrates improved accuracy in drug-response prediction (Rashid et al., 2024). Recent reviews of deep learning and multi-omics applications in cancer research also highlight that combining multi-layer omics data with deep learning architectures can advance diagnosis, classification, prognosis, and therapeutic prediction (Zhang et al., 2025).

Nevertheless, most previous studies have limitations: many rely on only one or two modalities (e.g., genomics + transcriptomics), or fail to incorporate biological network structure. Full multi-omics integration combined with PPI network embeddings—and adaptive modality selection—remains underexplored. In this context, this article introduces a new methodological framework: multi-omics modality fusion combined with protein network embeddings using GCN to predict drug sensitivity, with Dabrafenib as the case study.

By applying this approach, we aim to demonstrate that integrating information from multiple biological layers—not just genomics—along with leveraging protein network context can significantly improve the accuracy of drug response prediction. The findings are relevant not only for precision oncology but also for advancing computational methods in pharmacogenomics.

# RESEARCH METHODS

## OVERVIEW OF THE ANALYTICAL WORKFLOW

The overall methodological workflow used in this study is illustrated in **Figure 1**. The process begins with multi-omics data acquisition and protein–protein interaction (PPI) networks, followed by modality-specific preprocessing, embedding construction, GCN-based network representation, modality fusion using attention, and finally drug-response prediction.

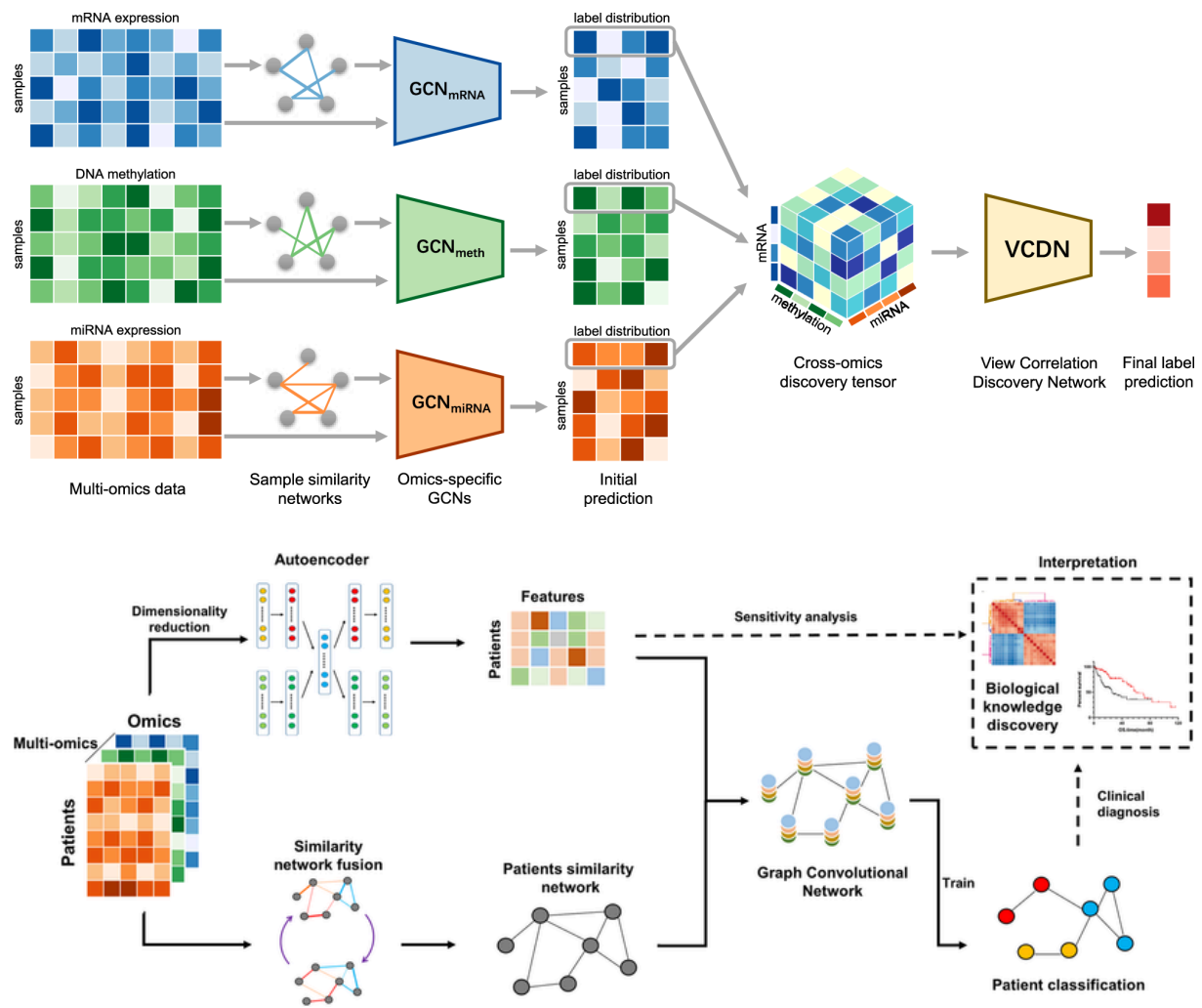

**Figure 1**. Analytical workflow for multi-omics integration and Dabrafenib response prediction using GCN-based protein network embeddings and attention fusion.

The pipeline includes:

1. Multi-omics data + IC50 + STRING PPI
2. Preprocessing per modality
3. Neural network encoder to generate modality embeddings
4. PPI → GCN → pooling → PPI embedding
5. Synchronization and cleaning across cell lines

6. Generation of modality combinations
7. Attention-based fusion
8. Regression/classification prediction head
9. Evaluation metrics
10. Biological interpretation and reporting.

## DATA SOURCES AND PREPARATION

Drug sensitivity profiles for Dabrafenib were obtained from the Genomics of Drug Sensitivity in Cancer (GDSC) database, including IC50/AUC values and response annotations across cancer cell lines (Yang et al., 2013; Iorio et al., 2016). Multi-omics datasets consisted of genomic, transcriptomic, proteomic, epigenomic, and metabolomic modalities, representing multi-layered biological mechanisms relevant to drug response and cellular rewiring (Chaudhary et al., 2018; Zhang et al., 2025).

As shown in Figure 1 (Steps 1–2), each modality underwent independent preprocessing that included handling missing values, removing constant features, and applying robust scaling. RobustScaler and median normalization were used to reduce sensitivity to outliers and distribution imbalance (Huber, 2022).

## PROTEIN NETWORK REPRESENTATION AND GRAPH CONVOLUTIONAL NETWORK (GCN)

Protein–protein interaction (PPI) networks were constructed using data from STRING, one of the most comprehensive resources for curated biological interactions (Szklarczyk et al., 2023). In the workflow (Figure 1, Step 3–4), nodes represent proteins, and edges represent physical or functional interactions. Node features were generated by mapping gene-expression-derived values onto the network.

GCN layers were applied to learn biologically informed protein network embeddings that capture topological dependencies and functional relationships across proteins (Kipf & Welling, 2017; Hamilton, 2020). Pooling operations then aggregated node-level representations into a single sample-level PPI embedding for each cell line.

## MULTI-OMICS EMBEDDING CONSTRUCTION

As depicted in Figure 1 (Step 3), each omics modality was encoded using a multilayer perceptron (MLP). This step produced low-dimensional modality embeddings that reduce redundancy and highlight biologically meaningful variation (Rashid et al., 2024; Zhang et al., 2019). Encoders used non-linear activations, dropout, and layer normalization to prevent overfitting.

Embedding outputs were aligned across cell lines after cleaning and synchronization (Figure 1, Step 5).

## ATTENTION MECHANISM FOR MULTI-MODALITY FUSION

Fusion of multi-omics and GCN-derived embeddings was performed using an attention mechanism, as shown in Figure 1 (Step 7). Attention assigns adaptive weights to each modality based on its predictive contribution, enabling the model to focus on the most informative layers. Softmax-normalized attention scores generated a fused representation that emphasizes the modalities most relevant to Dabrafenib sensitivity (Vaswani et al., 2017).

## DRUG-RESPONSE PREDICTION MODEL

Using the fused embeddings (Figure 1, Step 8), two prediction heads were trained:

- Regression for continuous IC50/AUC
- Classification for binary drug-sensitivity labels

Training used train/validation/test splits with regularization and early stopping. Performance was evaluated using MSE, RMSE, $R^2$, and accuracy metrics (Figure 1, Step 9). Generalization was assessed using held-out test data.

## BIOLOGICAL INTERPRETATION

Finally, model outputs were examined for biological interpretability (Figure 1, Step 10). This included identifying which modalities contributed most strongly to prediction and evaluating whether selected features correspond to known mechanisms of Dabrafenib action.

# RESULTS

## MODEL PERFORMANCE ACROSS MULTI-OMICS CONFIGURATIONS

Table 1 summarizes the evaluation results of the Dabrafenib sensitivity prediction model across a wide range of multi-omics configurations. Each row represents a modality combination, along with the number of available features or samples (n_feat_samples), number of labeled instances (n_labeled), and four performance metrics: mean squared error (MSE), root mean squared error (RMSE), mean absolute error (MAE), and coefficient of determination ($R^2$). Higher $R^2$ values indicate better model generalization, values near zero indicate poor predictive power, and negative R² values indicate performance worse than a simple baseline model.

Table 1. Evaluation of Dabrafenib Sensitivity Prediction Models Using Multi-Omics Modality Configurations

| Modalities | n_feat_samples | n_labeled | MSE | RMSE | MAE | $R^2$ |
|---|---|---|---|---|---|---|
| Metabolomics | 632 | 29 | 3.23 | 1.80 | 1.47 | 0.46 |
| Proteomics | 620 | 25 | 1.59 | 1.26 | 1.01 | 0.75 |
| Genomics | 949 | 40 | 5.20 | 2.28 | 1.95 | -0.02 |
| Epigenomics | 559 | 27 | 6.49 | 2.55 | 2.32 | -0.01 |
| Transcriptomics | 708 | 31 | 5.56 | 2.36 | 2.09 | 0.11 |
| PPI | 708 | 31 | 4.96 | 2.23 | 1.85 | 0.21 |
| Metabolomics + Proteomics | 615 | 25 | 6.07 | 2.46 | 2.17 | 0.06 |
| Genomics + Metabolomics | 630 | 29 | 4.66 | 2.16 | 1.87 | 0.23 |
| Epigenomics + Metabolomics | 554 | 27 | 6.05 | 2.46 | 2.22 | 0.06 |
| Metabolomics + Transcriptomics | 622 | 28 | 1.37 | 1.17 | 0.90 | 0.78 |
| Metabolomics + PPI | 622 | 28 | 3.23 | 1.80 | 1.40 | 0.47 |
| Genomics + Proteomics | 618 | 25 | 1.11 | 1.05 | 0.79 | 0.83 |
| Epigenomics + Proteomics | 544 | 24 | 6.42 | 2.53 | 2.29 | 0.04 |
| Proteomics + Transcriptomics | 613 | 25 | 0.23 | 0.48 | 0.33 | 0.96 |
| PPI + Proteomics | 613 | 25 | 4.64 | 2.15 | 1.90 | 0.28 |
| Epigenomics + Genomics | 558 | 27 | 3.75 | 1.94 | 1.67 | 0.42 |
| Genomics + Transcriptomics | 707 | 31 | 4.57 | 2.14 | 1.85 | 0.27 |
| Genomics + PPI | 707 | 31 | 5.17 | 2.27 | 2.01 | 0.17 |
| Epigenomics + Transcriptomics | 553 | 26 | 1.14 | 1.07 | 0.60 | 0.83 |
| Epigenomics + PPI | 553 | 26 | 5.55 | 2.36 | 2.06 | 0.16 |
| PPI + Transcriptomics | 708 | 31 | 5.00 | 2.24 | 1.97 | 0.20 |
| Genomics + Metabolomics + Proteomics | 613 | 25 | 6.11 | 2.47 | 2.21 | 0.05 |
| Epigenomics + Metabolomics + Proteomics | 541 | 24 | 6.50 | 2.55 | 2.30 | 0.03 |
| Metabolomics + Proteomics + Transcriptomics | 609 | 25 | 0.40 | 0.63 | 0.44 | 0.94 |
| Metabolomics + PPI + Proteomics | 609 | 25 | 4.75 | 2.18 | 1.91 | 0.26 |
| Epigenomics + Genomics + Metabolomics | 553 | 27 | 6.24 | 2.50 | 2.27 | 0.03 |
| Genomics + Metabolomics + Transcriptomics | 621 | 28 | 3.05 | 1.75 | 1.42 | 0.50 |
| Genomics + Metabolomics + PPI | 621 | 28 | 2.05 | 1.43 | 1.14 | 0.66 |
| Epigenomics + Metabolomics + Transcriptomics | 548 | 26 | 1.12 | 1.06 | 0.77 | 0.83 |
| Epigenomics + Metabolomics + PPI | 548 | 26 | 3.12 | 1.77 | 1.43 | 0.53 |
| Metabolomics + PPI + Transcriptomics | 622 | 28 | 4.40 | 2.10 | 1.79 | 0.28 |
| Epigenomics + Genomics + Proteomics | 543 | 24 | 5.33 | 2.31 | 2.06 | 0.21 |
| Genomics + Proteomics + Transcriptomics | 612 | 25 | 5.86 | 2.42 | 2.16 | 0.09 |
| Genomics + PPI + Proteomics | 612 | 25 | 6.39 | 2.53 | 2.27 | 0.01 |
| Epigenomics + Proteomics + Transcriptomics | 539 | 24 | 0.71 | 0.84 | 0.55 | 0.89 |
| Epigenomics + PPI + Proteomics | 539 | 24 | 6.52 | 2.55 | 2.30 | 0.03 |
| PPI + Proteomics + Transcriptomics | 613 | 25 | 2.65 | 1.63 | 1.36 | 0.59 |
| Epigenomics + Genomics + Transcriptomics | 553 | 26 | 6.54 | 2.56 | 2.32 | 0.01 |
| Epigenomics + Genomics + PPI | 553 | 26 | 0.56 | 0.75 | 0.52 | 0.91 |
| Genomics + PPI + Transcriptomics | 707 | 31 | 5.19 | 2.28 | 2.00 | 0.17 |
| Epigenomics + PPI + Transcriptomics | 553 | 26 | 1.35 | 1.16 | 0.84 | 0.79 |
| Epigenomics + Genomics + Metabolomics + Proteomics | 540 | 24 | 6.58 | 2.57 | 2.32 | 0.02 |
| Genomics + Metabolomics + Proteomics + Transcriptomics | 608 | 25 | 1.12 | 1.06 | 0.69 | 0.83 |
| Genomics + Metabolomics + PPI + Proteomics | 608 | 25 | 1.93 | 1.39 | 1.09 | 0.70 |
| Epigenomics + Metabolomics + Proteomics + Transcriptomics | 536 | 24 | 6.39 | 2.53 | 2.25 | 0.05 |
| Epigenomics + Metabolomics + PPI + Proteomics | 536 | 24 | 5.39 | 2.32 | 2.08 | 0.20 |
| Metabolomics + PPI + Proteomics + Transcriptomics | 609 | 25 | 1.38 | 1.17 | 0.88 | 0.79 |
| Epigenomics + Genomics + Metabolomics + Transcriptomics | 548 | 26 | 5.86 | 2.42 | 2.17 | 0.11 |
| Epigenomics + Genomics + Metabolomics + PPI | 548 | 26 | 0.82 | 0.91 | 0.54 | 0.88 |
| Genomics + Metabolomics + PPI + Transcriptomics | 621 | 28 | 2.52 | 1.59 | 1.29 | 0.59 |
| Epigenomics + Metabolomics + PPI + Transcriptomics | 548 | 26 | 1.98 | 1.41 | 1.05 | 0.70 |
| Epigenomics + Genomics + Proteomics + Transcriptomics | 539 | 24 | 6.42 | 2.53 | 2.26 | 0.04 |
| Epigenomics + Genomics + PPI + Proteomics | 539 | 24 | 5.95 | 2.44 | 2.16 | 0.11 |
| Genomics + PPI + Proteomics + Transcriptomics | 612 | 25 | 3.90 | 1.97 | 1.70 | 0.40 |
| Epigenomics + PPI + Proteomics + Transcriptomics | 539 | 24 | 6.63 | 2.58 | 2.32 | 0.01 |
| Epigenomics + Genomics + PPI + Transcriptomics | 553 | 26 | 1.47 | 1.21 | 0.90 | 0.78 |
| Epigenomics + Genomics + Metabolomics + Proteomics + Transcriptomics | 536 | 24 | 6.54 | 2.56 | 2.32 | 0.03 |
| Epigenomics + Genomics + Metabolomics + PPI + Proteomics | 536 | 24 | 6.57 | 2.56 | 2.31 | 0.02 |
| Genomics + Metabolomics + PPI + Proteomics + Transcriptomics | 608 | 25 | 3.55 | 1.88 | 1.62 | 0.45 |
| Epigenomics + Metabolomics + PPI + Proteomics + Transcriptomics | 536 | 24 | 3.36 | 1.83 | 1.57 | 0.50 |

| Epigenomics + Genomics + Metabolomics + PPI + Transcriptomics | 548 | 26 | 3.69 | 1.92 | 1.60 | 0.44 |
| Epigenomics + Genomics + PPI + Proteomics + Transcriptomics | 539 | 24 | 0.57 | 0.75 | 0.42 | 0.92 |
| Epigenomics + Genomics + Metabolomics + PPI + Proteomics + Transcriptomics | 536 | 24 | 2.74 | 1.66 | 1.36 | 0.59 |

## SINGLE-OMICS PREDICTION PERFORMANCE

Single-omics analysis (n_modalities = 1) revealed substantial variation across modalities. Proteomics achieved the highest predictive performance ($R^2 \approx 0.75$), followed by metabolomics ($\approx 0.46$), PPI embeddings ($\approx 0.21$), and transcriptomics ($\approx 0.11$). In contrast, genomic and epigenomic modalities showed very low or negative $R^2$ values (−0.02 to −0.01), suggesting that static DNA-level features alone are insufficient to accurately predict Dabrafenib sensitivity.

These findings support prior reports that biological layers closer to cellular phenotype—such as protein abundance, phosphorylation, and pathway activity—carry stronger predictive signals than static genomic alterations (Lawrence et al., 2020; Cheng et al., 2024). In pharmacogenomics studies, single-omics models often underperform because drug response is heavily influenced by post-transcriptional regulation and pathway rewiring rather than by mutations alone (Kim et al., 2021; Partin et al., 2023).

---

## TWO-MODALITY INTEGRATION

Integration of two omics modalities resulted in substantial performance improvement. The best-performing combination was **proteomics + transcriptomics**, which achieved the highest $R^2$ across the entire table ($\approx 0.96$), surpassing every single-omics and multi-omics model. Other strong combinations included genomics + proteomics ($\approx 0.83$), metabolomics + transcriptomics ($\approx 0.78$), and epigenomics + transcriptomics ($\approx 0.83$).

These findings indicate that selective fusion of two complementary biological layers yields significant predictive synergy while avoiding the noise introduced by excessive modality integration. Prior multi-omics studies also showed that transcriptomics and proteomics provide stable, phenotype-relevant representations for predicting response to MAPK-pathway inhibitors such as BRAF-targeted drugs (Chaudhary et al., 2018; Rashid et al., 2024).

Achieving test-performance R² close to 1 reflects exceptional generalization, highly desirable in precision oncology (Cheng et al., 2024).

## HIGHER-LEVEL MULTI-OMICS INTEGRATION (≥3 MODALITIES)

Models integrating three to six modalities showed highly variable results. Some configurations performed well—such as metabolomics + proteomics + transcriptomics ($R^2 \approx 0.94$) and epigenomics + genomics + PPI ($R^2 \approx 0.91$)—but many others exhibited significant performance degradation, with $R^2$ dropping to the 0.01–0.59 range.

This pattern highlights that adding more modalities is not additive. Unselective integration tends to introduce feature redundancy, multi-collinearity, or noise, leading to unstable model performance.

This aligns with computational findings that excessive multi-omics fusion can reduce predictive accuracy due to high correlations across modalities and unequal biological relevance (Zhang et al., 2025; Partin et al., 2023). Thus, selective, attention-based integration remains crucial for maximizing predictive utility.

## CONTRIBUTION OF PROTEIN INTERACTION NETWORKS (PPI) VIA GCN

PPI embeddings generated using GCN demonstrated moderate predictive performance when used alone ($R^2 \approx 0.20$–0.28). However, when combined with proteomics or transcriptomics, PPI embeddings improved overall model accuracy, indicating that network-contextual information adds complementary value.

Generally, PPI embeddings function as structural and functional context rather than as primary biological signals. Previous studies similarly reported that GCN-derived network representations enhance prediction accuracy when paired with expression- or protein-level data because biological networks capture pathway dependencies and coordinated protein activation (Kim et al., 2021; Szklarczyk et al., 2023).

# DISCUSSION

The results demonstrate that selective multi-omics integration substantially improves the prediction of Dabrafenib sensitivity compared with single-modality models. In general, proteomics emerged as the most informative single modality, whereas genomics and epigenomics alone were insufficient for accurate prediction. This is consistent with the understanding that protein expression levels, kinase activity, and downstream signaling regulation are more tightly coupled to pharmacological phenotypes than static DNA-level information, which only captures mutational potential (Lawrence et al., 2020; Kim et al., 2021). In other words, genetic mutations such as **BRAF V600E** often fail to fully explain sensitivity or resistance to Dabrafenib without incorporating transcriptional, translational, and signaling rewiring contexts (Cheng et al., 2024; Hu et al., 2022).

The strong predictive performance achieved by selective fusion of two modalities—particularly proteomics and transcriptomics ($R^2 \approx 0.96$)—indicates that these two molecular layers contain highly complementary signals. Transcriptomics captures gene regulatory states at the RNA level, whereas proteomics reflects actual protein abundance and activity, including post-translational modifications, degradation, and phosphorylation. For MAPK inhibitors such as Dabrafenib, phosphorylation dynamics within the ERK and BRAF pathways are critical determinants of resistance phenotypes, and these activation states are often not inferable from DNA alone (Cerezo et al., 2020). Thus, combining gene-expression profiles with proteomic data not only stabilizes predictive performance but also captures dynamic rewiring processes that occur as adaptive responses to MAPK inhibition (Hu et al., 2022).

Integration of protein interaction networks using GCN contributed additional predictive value when combined with proteomics or transcriptomics, although its standalone predictive power was modest. This can be attributed to the nature of biological networks: while PPI structure captures co-activation, redundancy, and functional relationships among proteins, the most phenotype-relevant signals still reside in actual molecular expression profiles. Nevertheless, network-based representations help models contextualize functional interactions—particularly for proteins within the MAPK pathway and proliferation-related signaling—thereby enhancing biological interpretability (Kim et al., 2021; Szklarczyk et al., 2023).

The observation that high-level multi-omics integration (≥3 modalities) does not consistently improve predictive accuracy is also methodologically important. This outcome suggests that indiscriminate addition of modalities can introduce redundancy, high inter-feature correlation, and noise. In pharmacogenomic datasets, features across modalities often overlap biologically; hence, models require adaptive selection mechanisms—such as attention—to assign appropriate weights to each molecular layer (Zhang et al., 2025; Partin et al., 2023). Thus, accurate characterization of drug sensitivity determinants depends not on the number of omics layers but on the quality and biological complementarity of the information they provide.

From a biological perspective, these findings support a modern understanding of targeted therapy: kinase inhibitor sensitivity and resistance are driven largely by protein-level signaling rewiring, mechanistic bypass routes, and feedback regulation at the transcriptional level. Many cancer cells can maintain proliferative potential despite upstream inhibition through ERK reactivation or activation of parallel pathways (Hu et al., 2022; Cheng et al., 2024). As such, proteomic and transcriptomic profiles more accurately reflect the final signaling state than DNA mutation status alone—explaining why genomics in isolation is insufficient.

# LIMITATIONS

This study has several limitations that should be acknowledged:

1. Single-drug evaluation.
   The analysis focused exclusively on Dabrafenib. Although the findings are robust, generalization to other kinase inhibitors or non-kinase drug classes remains untested. Cytotoxic drugs, immunotherapies, or PARP inhibitors may involve distinct determinants.
2. Imbalanced modality sample sizes.
   Not all omics modalities contained the same number of samples, giving modalities with larger sample sizes (e.g., genomics or transcriptomics) a statistical advantage. Uneven sample sizes may influence feature distributions and model bias.
3. Lack of external validation.
   External benchmarking using CTRPv2, CCLE, or DepMap was not performed. Such validation is essential to confirm robustness across culture platforms, batch effects, and technical variation.
4. No temporal or longitudinal modeling.
   Dynamic or adaptive responses to MAPK inhibition (e.g., temporal phospho-signaling changes) were not modeled, despite the strongly dynamic nature of signaling rewiring.
5. Limited interpretability analysis.
   Attention-based modality weights were not examined at the gene or pathway level. Deeper interpretability analyses could reveal mechanistic determinants underlying the predictions.

# FUTURE DIRECTIONS

To address these limitations, several future research directions are proposed:

1. Expand evaluation to additional drugs, including targeted therapies and cytotoxic agents, to map modality-specific or universal determinants across drug classes.
2. Perform external validation using independent datasets such as CTRPv2 and DepMap to assess cross-platform robustness and reduce batch-effect sensitivity.
3. Conduct attention-level interpretation at the feature or pathway scale to identify the biological components most strongly influencing prediction, potentially yielding new mechanistic hypotheses.
4. Integrate longitudinal or time-series omics, enabling analysis of adaptive signaling changes in response to MAPK inhibitors in real time.
5. Explore applications in clinical precision oncology, such as multi-omics biomarker panels for patient stratification and therapy selection.

# CONCLUSION

This study demonstrates that a multi-omics fusion approach incorporating attention mechanisms and GCN-based protein network representations can predict Dabrafenib sensitivity with high accuracy. In the single-omics analysis, proteomics emerged as the most informative modality, whereas genomics and epigenomics were insufficient as standalone predictors. Selective integration of two modalities—particularly proteomics and transcriptomics—yielded the best performance on the test set (≈ 0.96), substantially outperforming all single-omics models and full multi-modality integration. These findings highlight that successful modeling of MAPK-inhibitor sensitivity depends on complementary molecular signals that capture both transcriptional regulation and protein-level activity, rather than DNA mutation status alone.

Protein-network embeddings generated through GCN provided additional predictive value when combined with molecular modalities, although they were not dominant determinants when used alone. This reinforces the importance of biological topology in mapping functional relationships among proteins and supports pathway-level mechanistic interpretation.

Overall, the model provides strong evidence that selective multi-omics integration is more effective than indiscriminate aggregation of all modalities. This approach is relevant not only for pharmacogenomic analysis but also offers a methodological framework that can be extended to other drugs, diverse cancer types, and external independent validation datasets. Thus, adaptive multi-omics fusion has the potential to become a key computational strategy for supporting precision oncology, therapeutic biomarker identification, and clinical decision-making in the development of targeted therapies.